\documentclass[aps,showpacs,prb,floatfix,twocolumn]{revtex4}
\usepackage{amsmath,amssymb,graphicx,bm,epsfig}

\input{tcilatex}

\begin{document}

\title{Program LMTART\ for Electronic Structure Calculations}
\author{S. Y. Savrasov}
\affiliation{Department of Physics, New Jersey Institute of Technology, Newark, NJ 07102,
USA}

\begin{abstract}
A computer program LMTART for electronic structure calculations using full
potential linear muffin--tin orbital method is described.
\end{abstract}

\maketitle

\section{DESCRIPTION}

Full--potential linear--muffin--tin--orbital (Ref. \cite{OKA}) (FP--LMTO)
program LMTART is designed to perform band structure, total energy and force
calculations of solids using methods of density functional theory (DFT)
(Ref. \cite{GAS}). The development of LMTART\ has been initiated in 1986 in
P.N. Lebedev Physical Institute. It was further extensively developed in
Max--Planck Institut f\"{u}r Festk\"{o}rperforschung, and its most recent
contributions have been done in Departments of Physics of Rutgers University
and of New Jersey Institute of Technology.

LMTART\ performs electronic structure calculations using popular local
density approximation (LDA) and generalized gradient approximation (GGA) of
density functional theory \cite{GAS}. To deal with strongly correlated
systems it also implements LDA+U method \cite{ANIS} and will include the
capability to study materials using spectral density functional approach and
dynamical mean field theory (DMFT) \cite{SDFT} in the nearest future. For
determining equilibrium crystal structure and phonon spectra LMTART\
computes total energies, forces \cite{FORCE} as well as full wavevector
dependent lattice dynamics \cite{LRPRB} using linear response theory. For
studying optical spectral functions LMTART\ evaluates dipole matrix elements
between various electronic states and can determine optical conductivity of
materials.

The code implements atom centered local muffin--tin orbitals as a basis for
representing one--electron wave functions in both bare and screened
(tight--binding) representations with general choice of tail energies $%
\kappa ^{2}.$ This allows to increase accuracy in the calculations by using
multiple--$\kappa $ LMTO expansions. Low lying core electronic states can be
resolved as bands in separate energy panels. For heavy elements, spin--orbit
coupling matrix elements can be taken into account using a variational
procedure proposed by Andersen \cite{OKA}. Full three dimensional treatment
of magnetization in relativistic calculations including LDA+U is implemented
to study problems related to magnetic anisotropy and non--collinear
magnetism. Effect of finite temperature can be modelled by introducing Fermi
statistics for the one--electron states. Additionally, LMTART\ can determine
tight--binding fits to the energy bands by extracting hopping integrals
between various orbitals using tight--binding LMTO representation.

LMTART\ is written on FORTRAN 90 and uses dynamical memory scheme. No
additional recompilation of the code is required when changing numbers of
atoms, spins, plane waves, etc. There are two basic regimes for working with
this program: the self--consistent charge density calculation and
calculation of physical properties such as electronic structure, optical
properties, etc.. The simplest input to LMTART involves only atomic charges
of the atoms as well as crystal structure.

LMTART\ works with two different approximations related to a shape of the
potential: (i) {atomic sphere approximation (ASA)} and plane wave expansion
(PLW). ASA uses overlapping atomic spheres, where the potential is expanded
in spherical harmonics inside the spheres, but any contribution from the
interstitial region is neglected. Such method is fast and provides
reasonably good energy bands. However it is not sufficiently accurate to
deal with distortions and phonons. {PLW }is a full potential approximation
which uses non--overlapping muffin--tin spheres, where the potential is
represented via spherical harmonics expansions, and the interstitial region
where the potential is expanded in plane waves. The full potential regime
provides the best accuracy at the price of increasing computational time.
Finally, LMTART can be run in a special tight--binding regime by supplying
hopping integrals between various orbitals.

Two extensions of LMTART are available as separate programs. One is a
full--potential linear--response linear--muffin--tin--orbital package LMTO
PHONONS which is designed to perform linear--response calculations of the
phonon spectra for arbitrary wave vectors \textbf{q}. Main features of this
code include: (i) Computations of first--order changes in the charge density
and the potential due to displacements of nuclei for arbitrary wave vectors 
\textbf{q}. (ii) Calculations of the dynamical matrix and the phonon
spectra. (iii) Calculations of electron--phonon interactions, Eliashberg
spectral functions $\alpha ^{2}F(\omega )$ and transport properties such as
electrical and thermal resistivities. The description of the method and
complete set of references can be found in Ref. \cite{LRPRB}. Numerous
applications are given in Ref. \cite{APPL}.

Also available is an extension of LMTART to compute dynamical
susceptibilities and magnon spectra This is a full--potential
linear--response linear--muffin--tin--orbital package of programs LMTO
MAGNONS designed for these purposes. As a by--product, spin wave spectra and
their lifetimes are accessible as peaks in calculated imaginary spin
susceptibility. A short description of the method can be found in Ref. \cite%
{MAG}.

Software MINDLab which runs under Microsoft Windows operating systems can be
used to set up input files and analyze output files of LMTART. It can also
be used to run the\ code without learning its extensive input options. This
software is necessary to visualize all data withdrawn from LMTART.
Currently, MINDLab can perform the following tasks (i) Crystal group
calculations. By setting atomic positions MINDLab shows crystal group
operations found for given atomic configuration; (ii) Fat bands
calculations. Fully colored visualization of separated orbital characters on
top of the band structure can be performed by mouse click operations. (iii)
Density of states calculations. MINDLab computes and visualizes densities of
states, total and orbital resolved. (iv) Extractions of hopping integrals
for tight--binding fits can be performed quickly with MINDLab. (v) Optical
properties calculations. Dielectric functions $\epsilon _{1}(\omega )$, $%
\epsilon _{2}(\omega )$, and electron energy loss spectra can be computed
and visualized by MINDLab. (vi) Visualizations of charge densities, full
potentials, and Fermi surfaces in both two and three dimensions using OpenGL
graphical library implemented within MINDLab (viii) Three dimensional
visualization of crystal structures and their distortions. (ix) Full three
dimensional visualization of vector fields including velocity fields on top
of the Fermi surfaces and magnetizations.(x) Simplified input for studying
correlated electronic systems which currently implements LDA+U method and
will include LDA+DMFT technique in the nearest future.

MINDLab works with LMTART\ as with an executable file which reads input
data, performs band structure calculation, and stores output files. MINDLab
controls this process, it prepares input for LMTART using dialog windows,
and executes LMTART program as a separate thread. When the calculation is
finished, LMTART notifies the MINDLab, and the output files can be
visualized by simple mouse click operations. While running LMTART for
complicated compounds can be rather slow (it depends how many atoms per unit
cell is selected), MINDLab can be used to prepare the input, after which the
files can be copied to a more powerful computer where the LMTART can be
executed much faster. After run is performed, the output can be copied back
to the PC and analyzed quickly by MINDLab. All input/output files of LMTART
are formatted ASCII files, and therefore system independent.

\section{APPLICATIONS}

Several most recent applications of LMTART can be found in Refs. \cite%
{SDFT,APPL,MAG,MgCNi3,MAE}. They include calculations of phonons and
electron--phonon interactions in novel superconductors, such as MgB$_{2},$%
LiBC, MgCNi$_{3},$ studies of phonons in strongly correlated systems, e.g.
NiO, Pu, computations of magnetic anisotropy energies of ferromagnets Fe,
Co, Ni, CrO$_{2}\,$\ evaluations of dynamical spin susceptibilities and spin
wave spectra, and so on.

Here we illustrate one of the most recent calculation \cite{MgCNi3} of
phonon spectrum in MgCNi$_{3}$ using a linear response regime of LMTART. The
calculated phonon dispersions along major high symmetry lines of the cubic
Brillouin zone are given on Fig. 1. The frequencies are seen to be span up
to 1000\ K, with some of the modes showing significant dispersion. In
general, we distinguished three panels where the top three branches around
900 K are carbon based, the middle three branches around 600 K are Mg based
and 9 lower branches are all Ni based.

\begin{figure}[tbh]
\includegraphics*[height=2.8in]{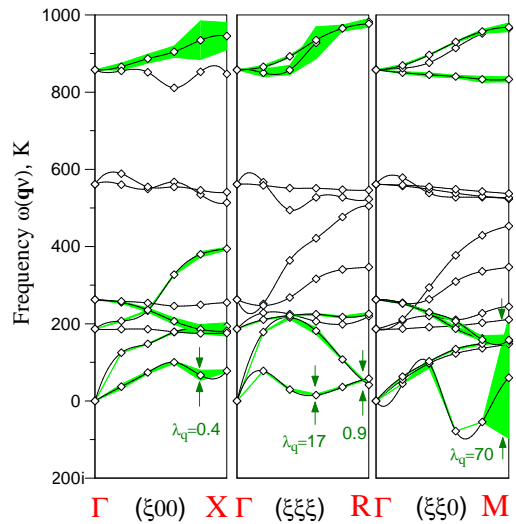}
\caption{Calculated phonon spectrum of MgCNi$_{3}$ along major symmetry
directions of the Brillouin zone of the cubic lattice using density
functional linear response method. Some curves are widened proportionally to
the phonon linewidths. }
\label{fig:Fig1}
\end{figure}

We have discovered a striking feature of this phonon spectrum connected to
the presence of a low--frequency acoustic mode which is very soft and is
even seen to be unstable along some points in $(\xi \xi 0$) direction in the
Brillouin zone. This mode is essentially Ni based and corresponds to
perpendicular movements of two Ni atoms towards octahedral interstitials of
the perovskite structure. The latter is made of each of the four Ni atoms
and two Mg atoms.

The persistence of the instability which does not occur for any of the
high--symmetry point needs a non--trivial frozen--phonon analysis. As our
polarization vectors prompt that the largest anharmonicity is expected for
the $\Gamma XM$ plane ($q_{z}=0$) of the Brillouin zone of the cubic
lattice, we have performed three such calculations for the points $a=$($%
\frac{1}{4}\frac{1}{4}0)\frac{2\pi }{a}$, $b=$($\frac{1}{2}\frac{1}{4}0)%
\frac{2\pi }{a}$, and $M.$ The results of these calculations reveal
essentially anharmonic interatomic Ni potentials. A shallow double well with
a depth of the order of 20--40 K and the curvature at the equilibrium of the
order of 40--50i K exists for the $a$ and $b$ points which becomes vanishing
at the M point. We have found such a behavior by both total energy and force
calculations for the supercells of 20 atoms ($a$ and $b$ points) and 10
atoms ($M$ point). Clearly, such a small depth on the temperature scale
indicates that the distortions are dynamical and zero point motions would
prevent the appearance of the static long--range order. Complete discussion
of these calculations as well as the comparisons with existing experiments
can be found in Ref. \cite{MgCNi3}.

Fig.1 also shows the calculated phonon linewidths $\gamma _{\mathbf{q}\nu }$%
by widening some representative phonon dispersion curves $\omega _{\mathbf{q}%
\nu }$ proportionally to $\gamma _{\mathbf{q}\nu }$. In particular, we found
that some phonons have rather large linewidths. This, for example holds, for
all carbon based higher lying vibrational modes. The strength of the
coupling, $\lambda _{\mathbf{q}\nu },$ for each mode can be obtained by
dividing $\gamma _{\mathbf{q}\nu }$ by $\pi N(\epsilon _{F})\omega _{\mathbf{%
q}\nu }^{2},$where $N(\epsilon _{F})$ is the density of states at the Fermi
level equal to 5.3 st./[eV*cell] in our calculation. Due to large $\omega _{%
\mathbf{q}\nu }^{2},$ this unfortunately results in strongly suppressed
coupling for all carbon modes which would favor high critical temperatures.
The coupling, however, is relatively strong for the Ni based modes. For
example, we can find $\lambda $'s of the order of 1--3 for the Ni based
optical phonons around 250\ K. Our resulting value of electron--phonon
coupling constant $\lambda $ is found 1.51. This assumes that MgCNi$_{3}$ is
a strongly coupled electron--phonon superconductor.

\section{DOWNLOADS}

All LMTART\ based programs described above are available for downloading at
the following URL: \emph{http://www.physics.njit.edu/\symbol{126}mindlab}.
Each program consists of the source code, manual, and example files. An
extensive database of electronic structures for different materials is also
available at this HTTP URL.

\section{\textit{Acknowledgements}}

Part of the development has been done in collaboration with my brother Dr.
Dmitrij Savrasov. Special thanks to Prof. Ole Andersen and Dr. Ove Jepsen
who are my LMTO teachers. I also greatly acknowledge Dr. Andrej Postnikov
who has initiated writing of all manuals.

The development of these programs is currently supported by the NSF DMR
Grants No. 02382188, 0342290,\ US DOE division of Basic Energy Sciences
Grant No. DE-FG02-99ER45761, and by the Computational Material Science
network operated by US\ DOE.

\end{document}